# Revealing the Global Linguistic and Geographical Disparities of Public Awareness to Covid-19 Outbreak through Social Media

Binbin Lin[a], Lei Zou[a]*, Nick Duffield[b], Ali Mostafavi[c], Heng Cai[a], Bing Zhou[a], Jian Tao[d], Mingzheng Yang[a], Debayan Mandal[a], Joynal Abedin[a]

[a]*Department of Geography, Texas A&M University, College Station, TX, USA*

[b]*Department of Electrical & Computer Engineering, Texas A&M University, College Station, TX, USA*

[c]*Department of Civil & Environmental Engineering, Texas A&M University, College Station, TX, USA*

[d]*Department of Visualization, Texas A&M University, College Station, TX, USA*

*\*Email: lzou@tamu.edu*

# Revealing the Global Linguistic and Geographical Disparities of Public Awareness to Covid-19 Outbreak through Social Media

The Covid-19 has presented an unprecedented challenge to public health worldwide. However, residents in different countries showed diverse levels of Covid-19 awareness during the outbreak and suffered from uneven health impacts. This study analyzed the global Twitter data from January 1st to June 30th, 2020, seeking to answer two research questions. What are the linguistic and geographical disparities of public awareness in the Covid-19 outbreak period reflected on social media? Can the changing pandemic awareness predict the Covid-19 outbreak? We established a Twitter data mining framework calculating the Ratio index to quantify and track the awareness. The lag correlations between awareness and health impacts were examined at global and country levels. Results show that users presenting the highest Covid-19 awareness were mainly those tweeting in the official languages of India and Bangladesh. Asian countries showed more significant disparities in awareness than European countries, and awareness in the eastern part of Europe was higher than in central Europe. Finally, the Ratio index could accurately predict global mortality rate, global case fatality ratio, and country-level mortality rate, with 21-30, 35-42, and 17 leading days, respectively. This study yields timely insights into social media use in understanding human behaviors for public health research.

## 1. Introduction

The outbreak of the novel coronavirus, known as Covid-19, has profoundly impacted human society. In 2020, Covid-19 had infected more than 83.48 million people and caused nearly 1.82 million deaths in 191 countries and regions (Dong, Du, and Gardner 2020). Since the Covid-19 outbreak, governments worldwide have implemented several measures requiring or suggesting residents wear masks, keep social distancing, or stay at home to control the spread of the coronavirus. However, residents in different countries showed diverse levels of awareness of Covid-19 and relevant policies during the outbreak and suffered from uneven health impacts, including unequal infection, fatality, and recovery rates (Gollust et al. 2020; McCaffery et al. 2020; Hu et al. 2020; Saad, Hassan, and Zaffar 2020). Whether the changes and disparities in public awareness led to different responding behaviors and thus affected the pandemic's health impacts is unknown and needs to be investigated.

However, continuous long-term and near-real-time data describing disparities in public awareness and responses cannot be obtained through traditional survey methods, especially during the pandemic. With the development of Web 2.0 and GNSS-enabled portable devices, social media platforms, e.g., Twitter, Facebook, and Instagram have become increasingly popular worldwide for sharing feelings and discussing 'what's happening'. Data collected from such platforms provide an emerging channel to observe real-time human responses to different topics and events (Zou et al. 2018). During Covid-19, many people's social lives have shifted from in-person to online to stay connected while maintaining social distancing, and they spent more time sharing their experiences, concerns, and feelings toward topics relevant to Covid-19 on social media (Alqurashi, Alhindi, and Alanazi 2020; Chen, Lerman, and Ferrara 2020; Lopez, Vasu, and Gallemore 2020). As a result, the extensive social media data generated during the pandemic offers an innovative opportunity to observe the public reactions to Covid-19 in near real-time.

Nevertheless, studying human behaviors at different locations during the pandemic and drawing scientific conclusions from social media data are challenging. First, social media data contain a sheer amount of noisy information irrelevant to Covid-19. It is difficult to accurately collect pandemic-related messages from the big social media database. Second, many social media platforms support multiple languages, and identifying and analyzing coronavirus-related messages of different languages necessitates advanced natural language processing (NLP) models (Lopez, Vasu, and Gallemore 2020). Third, social media users are unevenly distributed across space, and only a small proportion of the generated data (around 1-2%) contains precise locations (Graham, Hale, and Gaffney 2013). Social media data need to be associated with geographic contexts and normalized through preprocessing to enable spatial and temporal analytics.

This study analyzed the global Twitter data, referred to as tweets, from January 1$^{st}$ to June 30$^{th}$, 2020, when Covid-19 developed from a regional epidemic disease to a pandemic causing a global health crisis. The overarching research questions are: What are the linguistic and geographical disparities in public awareness of the Covid-19 outbreak on social media? Can the changing public awareness on social media predict the pandemic's outbreak? To address the research questions, three objectives are proposed and achieved: (1) to establish a social media data mining framework tracking the public awareness of Covid-19 by languages and regions; (2) to quantify disparities of awareness toward the pandemic at multiple spatial and temporal scales; and (3) to examine the lag correlations between Covid-19 awareness and health impacts globally and regionally. One hypothesis is tested: social media-derived public awareness can predict the pandemic's outbreak at both global and regional scales. The results can inform governments to mitigate risks from current and future epidemics through social media data analysis.

## 2. Background

### 2.1 Social Media Data Mining for Public Health

Social media data contain abundant attributes, e.g., time, geographical information,

contents, and relationships among users, providing a brand-new perspective for understanding human behaviors in spatial, temporal, contextual, and network dimensions. Since the emergence of social media, researcher have applied data collected from such platforms to address multiple health-related issues, e.g., obesity, depression, insomnia, and air pollution (Choudhury et al. 2013; Gore, Diallo, and Padilla 2015; McIver et al. 2015; Chen et al. 2017; Sun et al. 2018; Gao et al. 2020).

Sun et al. (2018) collected the obesity data from the Gallup Healthways Wellbeing Survey and the U.S. Centers for Disease Control and Prevention (CDC) and 41 million tweets from 110 major cities in the USA during 2012-2013. They proposed an obesity estimation method via monitoring users' dietary habits, physical activities, emotions, and self-consciousness on Twitter, demonstrating that user activities on online social networks could help evaluate the obesity rate in urban areas. Chen et al. (2017) developed a forecasting model for smog-related hazards by creating a Smog Severity Index and social media diffusion factor based on microblog data. The model-predicted smog severity was consistent with the measured values from ground and satellite sensors, implying the potential to monitor health-related disaster threats and impacts through social media platforms. McIver and others (2015) identified a list of insomniacs based on their Twitter messages and investigated their Twitter use behaviors. The results show that insomniacs had fewer followers, expressed lower sentiments, and were less active on social networks compared with other users. Social media data mining also has a great potential in addressing mental health issues. A study finds that integrating social media users' activities, e.g., social engagement, linguistic styles, networks, and emotions, could characterize and forecast individual-level depression (Choudhury et al. 2013).

In addition to the above case studies focusing on specific health diseases, Paul and Dredze (2014) built a topic model called Ailment Topic Aspect Model (ATAM) to automatically discover health-relevant topics on Twitter without human supervision or a priori knowledge. Culotta (2014) created 160 indexes based on Twitter data and performed regression analysis to predict the county-level statistics of 27 types of health conditions (e.g., obesity, teen births, and diabetes) in 100 counties in the United States. The results show that the predictive models incorporating Twitter-derived variables have higher accuracy for surveying county-level health conditions compared to the models based on traditional questionnaires.

**2.2 Social Media and Covid-19**

Since the beginning of 2020, researchers from different fields have made enormous efforts to collect and mine social media data during Covid-19 to understand human behaviors during the pandemic and combat it. For instance, several studies collected pandemic-related social media data and shared them through open-source archives (Alqurashi, Alhindi, and Alanazi 2020; Chen, Lerman, and Ferrara 2020; Lopez, Vasu, and Gallemore 2020). Rufai and Bunce (2020) pointed out that Twitter empowered world leaders to exchange Covid-19 information with citizens rapidly. La et al. (2020) analyzed the official Covid-19 news from online newspapers and concluded that immediate policies and supportive public responses are the key epidemic control measures. Sentiment analysis and epidemic-related topics surveillance are essential for understanding social activities during the pandemic (Chen et al. 2020; Lwin et al. 2020; Nemes and Kiss 2020; Zhu et al. 2020). Leveraging social media data, worldwide trends of four emotions, fear, anger, sadness, and joy, were examined in Lwin et al. (2020). Their results show that public feelings shifted strongly from fear to anger from January 28$^{th}$ to April 9$^{th}$, 2020.

Predicting the regional epidemic outbreak from social media activities has also been proved feasible in several investigations (Jahanbin and Rahmanian 2020; Li et al. 2020; Qin et al. 2020). Specifically, Li et al. (2020) found that the Covid-19 discussion peak on social media occurred 10-14 days earlier than the peak of daily incidences in China. Online social networks are effective platforms for disseminating rumors or conspiracy theories (Allington et al. 2020; Gruzd and Mai 2020; Tasnim, Hossain, and Mazumder 2020). Shahsavari et al. (2020) investigated the narrative frameworks of fabricating and broadcasting Covid-19 conspiracies to monitor those messages near real-time.

Although the existing studies generate innovative methods and valuable knowledge in using social media for Covid-19 research, some limitations exist. For example, Covid-19 related social media data analysis across regions is needed to address the inequalities of public awareness and health impacts. Examining the long-term Covid-19 social media activities is also critical because Covid-19 is a long-standing epidemic and short-term analysis is unable to capture how the public perception of Covid-19 changes temporally.

## 3. Data and Methods

### 3.1 Twitter Data

This research collected social media data from Twitter, one of the most popular social media and network platforms with about 340 million users in more than 150 countries (Ahlgren 2021). Approximately 500 million tweets were published per day in various languages in 2020 (Ahlgren 2021), enabling the tracking of long-term global awareness of Covid-19. We collected Twitter data from January 1$^{st}$ to June 30$^{th}$, 2020 from Internet Archive (https://archive.org/), an online free data library that stores around 1 percent of the whole Twitter database. Tweets are encoded in JavaScript Object Notation (JSON) format and represented as a list of name-value pairs. Figure 1 shows the Twitter data collection and preprocessing workflow. Three attributes were used in the subsequent analysis, created_at, text, and lang, representing each tweet's timestamp, content, and language, respectively.

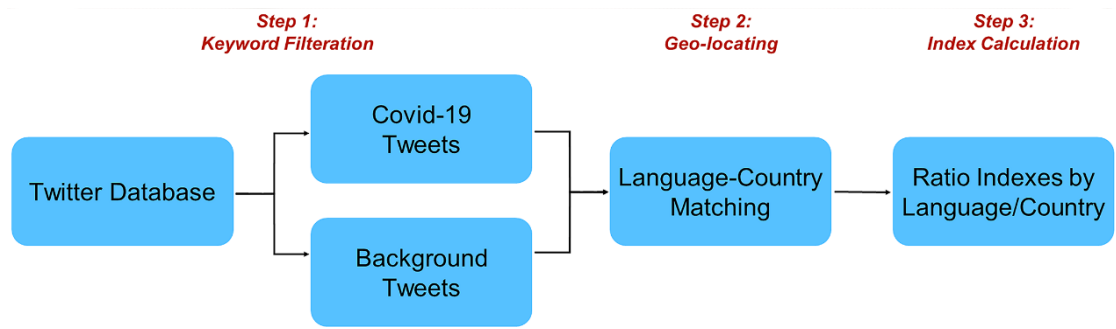

Figure 1. Workflow of Twitter data collection and preprocessing.

The first step was identifying Covid-19-related tweets. We selected eight keywords, i.e., covid, virus, corona, ncov, n95, pandemic, pneumonia, and quarantine, based on an overview of previous literature to search for the Covid-19 relevant tweets (Alqurashi, Alhindi, and Alanazi 2020; Chen, Lerman, and Ferrara 2020; Qin et al. 2020). Considering that Twitter supports multiple languages, we translated the eight keywords into different languages by the Google Translate Application Programming Interface (API) to collect the global Covid-19 discussion on Twitter. A total of 65 distinct languages (including English) were detected from the collected Twitter data, and the Google Translate API supports 61 of them. Any tweet containing at least one of the eight keywords in one of the 61 languages was identified as Covid-19-related.

The second step was to determine the location where each tweet was sent from. There are three common metadata sources to geo-reference tweets - geotagged location, user profile address, and content mentioned place. However, each source has its limitations for this study. First, less than 2% of the Twitter data contains the geotagged locations (Zou et al. 2018), and the proportion of geotagged Covid-19 tweets could be even less, which makes over 98% of the tweets unusable. Due to the low use of precision geotagging and the increasing concern for users' privacy, Twitter has gradually removed the function of attaching point coordinates to tweets since June 2019. Second, a high percentage of tweets (40% to 60%) can be linked to county-level locations using user profile addresses. However, it requires the use of geocoding services (e.g., Google or OpenStreetMap Geocoding API) for toponym resolution, which is a time-consuming and computationally intensive task when processing a large amount of data and introduces geocoding uncertainty (Zou et al. 2019; Wang et al. 2021). Further, deriving locations from the tweet content mentioned places is unreliable because places mentioned in tweet contents (tweet about) do not necessarily reflect the locations of Twitter users (tweet from).

Therefore, this study leveraged an alternative metadata source to associate each tweet with a country or a region: tweet language, which Twitter automatically detected and provided in the collected tweets. Users in each country are more likely to post messages on Twitter using their country's official/primary language (Mocanu et al. 2013; Zola, Cortez, and Carpita 2019). Thus, matching the tweet language and the official/primary language of each country could indicate Twitter users' native locations and cultural backgrounds. For instance, a tweet written in Thai is considered sending from Thailand since it is the only country that uses Thai as its official language. Among the 61 selected languages, 29 can be paired to a single country (Figure 2) based on the language-country matching list in Table A1. This language-country matching method has been applied in a previous investigation with an overall accuracy of 53% (Zubiaga et al. 2017). We validated the language-country matching approach by calculating the accuracy of language-detected countries in geotagged tweets. The results are summarized in section 4.1.

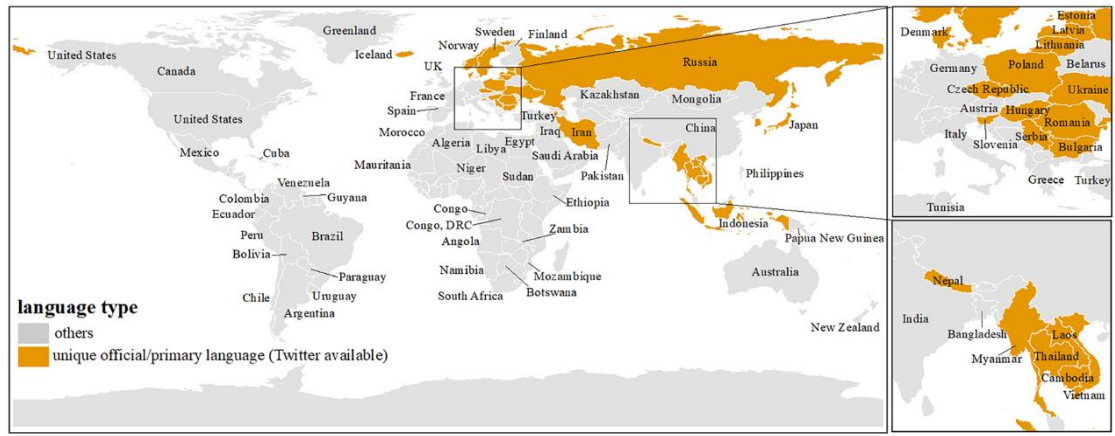

Figure 2. Language type of each country.

Third, we computed the Ratio index using Equation 1 by languages and at multiple spatial-temporal scales to compare the levels of public awareness to Covid-19 in different linguistic cultures, geographical regions, and pandemic phases (Zou et al. 2018). The Ratio index values range from 0 to 1. A small Ratio index value means a low level of public awareness about Covid-19 on Twitter and vice versa.

$$\text{Ratio} = \frac{\text{Number of Covid-19 related tweets}}{\text{Number of all tweets}} \qquad (1)$$

**3.2 Covid-19 Health Impact Indicators**

We chose three indicators - case fatality ratio, case rate, and mortality rate - to represent Covid-19's public health impacts. Case fatality ratio is defined as the proportion of people who died from Covid-19 among individuals diagnosed with Covid-19 (Equation 2). Case/mortality rate are calculated as the number of confirmed cases/deaths divided by the total population and normalized by 10 million (Equations 3&4). Cumulative data documenting global Covid-19 cases and deaths from January 22$^{th}$ to June 30$^{th}$, 2020, were obtained from the publicly available database in the

Center for Systems Science and Engineering (CSSE) at Johns Hopkins University (https://coronavirus.jhu.edu/map.html) (Dong, Du, and Gardner 2020). The population data were collected from the United Nations' 2020 mid-year population estimation (http://data.un.org/). The values of case fatality ratio, case rate, and mortality rate range from 0-100%, 0-$10^7$, and 0-$10^7$ respectively, and the large value of each index means a more severe health impact by Covid-19. These three indicators can be computed or aggregated by different areas (e.g., global, continental, and country) within a defined time period (e.g., daily, monthly, and yearly).

$$\text{case fatality ratio} = \frac{\text{\# Covid-19 deaths}}{\text{\# Covid-19 cases}} \tag{2}$$

$$\text{case rate} = \frac{\text{\# Covid-19 cases}}{\text{Population}} \times 10^7 \tag{3}$$

$$\text{mortality rate} = \frac{\text{\# Covid-19 deaths}}{\text{Population}} \times 10^7 \tag{4}$$

**3.3 Methods of Analysis**

The public awareness was aggregated and evaluated by different languages and at multiple spatial-temporal scales to examine its relationship with the Covid-19's health impact indicators. First, we calculated the daily Ratio index of tweets in all languages globally to indicate the general trend of public awareness toward Covid-19 on Twitter. The global daily case rate, mortality rate, and case fatality ratio were calculated to represent the general trend of health impacts caused by Covid-19. To examine if significant associations and lag effects exist between public awareness and the Covid-19 outbreak worldwide, the lag Pearson correlations between global Ratio index and the three Covid-19 health impact indicators were calculated using Equations 5-7.

$$\text{lag Pearson correlation}(X_t, Y) = \frac{Cov(X_t, Y)}{\sqrt{Var[X_t]Var[Y]}} \quad (5)$$

$$Cov(X_t, Y) = \frac{1}{n-1} \sum_{i=1}^{n} (X_{ti} - \bar{X}_t)(Y_i - \bar{Y}) \quad (6)$$

$$Var[X_t]Var[Y] = \frac{1}{(n-1)^2} \sum_{i=1}^{n} (X_{ti} - \bar{X}_t)^2 \sum_{i=1}^{n} (Y_i - \bar{Y})^2 \quad (7)$$

In Equation 5, $X_t$ denotes the time series of health impact indicators moved to the left by t days, and $Y$ represents the time sequences of Ratio index. $Cov(X_t, Y)$ is the covariance of $X_t$ and $Y$, calculated according to Equation 6. $Var[X_t]$ and $Var[Y]$ are the variances of $X_t$ and $Y$, and their product is calculated utilizing Equation 7.

Second, the Ratio for the half year and the monthly Ratio indexes by language were calculated to show the general differences and temporal variations of Covid-19 awareness among Twitter users with diverse cultural backgrounds. Third, we focused on countries having unique official/primary languages and generated maps of half-year and monthly Ratio indexes by country to reveal the spatial-temporal disparities of Covid-19 awareness. Finally, the lag correlations between Ratio index and Covid-19 health impact indicators were tested at the country level to examine the predictability of public awareness changes on the regional Covid-19 outbreak.

## 4. Results

This study identified a total of 10,339,830 (2.12%) out of over 488 million tweets as Covid-19-related (Table 1). Among all tweets, 218,086 (0.04%) have geotags, and the amount of geotagged Covid-19 tweets is 4,467 (0.0000091%). More than 450 million (92.26%) tweets have language (lang) attribute values and 10,184,313 (2.08%) of them are Covid-19 tweets. Within geotagged tweets, 53,702 can be matched to a single country by language. We compared the agreement of geotag-derived and tweet language inferred country information in the collected tweets to validate the language-country matching method. The overall accuracy is 77%, indicating that this method is reliable to geo-locate tweets for country-level analysis.

Table 1. Summary statistics for the collected tweets

| Twitter Data | Count | Percentage of Tweets (%) |
| --- | --- | --- |
| All Tweets | 488,486,181 | 100.00 |
| Covid-19 Tweets | 10,339,830 | 2.12 |
| Geotagged Tweets | 218,086 | 0.04 |
| Geotagged Covid-19 Tweets | 4,467 | $9.1*10^{-6}$ |
| Tweets with Language Attributes | 450,664,878 | 92.26 |
| Covid-19 Tweets with Language Attributes | 10,184,313 | 2.08 |

**4.1 Global Temporal Trends**

Figure 3 shows the global temporal patterns of the daily Ratio index and three health impact indicators from January 1st to June 30th, 2020, which reflect the general trend of worldwide public awareness and health impact of Covid-19 during its outbreak. Using the date when the World Health Organization (WHO) announced Covid-19 a pandemic as the breaking point, we defined January 1st to March 11th as the pre-pandemic period and March 12th to June 30th as the pandemic-outbreak period.

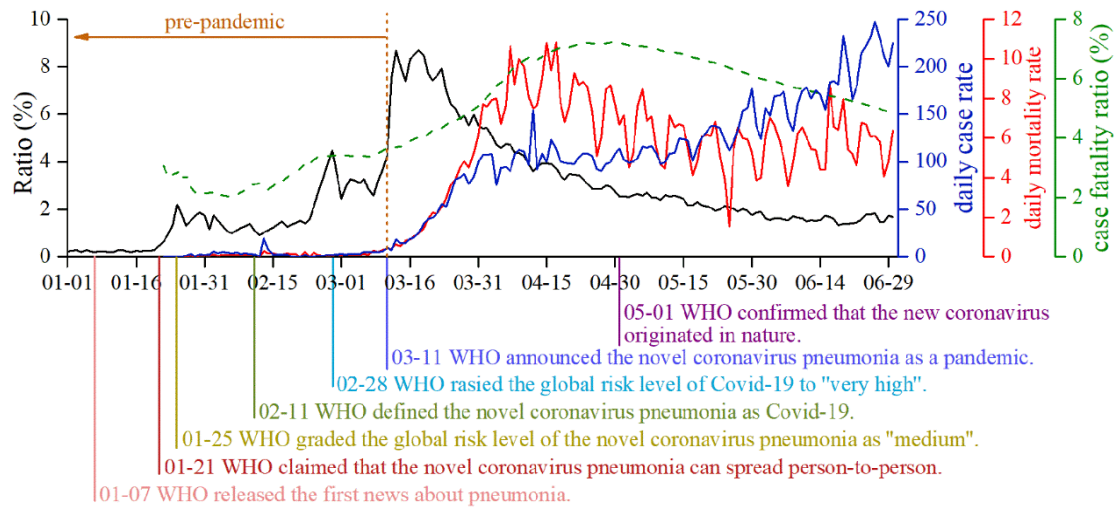

Figure 3. Global temporal trends of the Ratio Index and health impact indicators of Covid-19.

The daily Ratio index (black line) ranged from 0.20% to 8.71%, with an average value of 2.72%. The high values were found during March 12th to 23rd, and the maximum value was on March 18th, a week after entering the pandemic-outbreak period. During January 1st to 20th, very few tweets discussed Covid-19, even though it had been reported to spread in Wuhan, China. Since January 21st, when WHO claimed that the novel coronavirus pneumonia could spread from person to person, the Ratio index followed a fluctuating ascending trend with two peaks in the pre-pandemic period. One peak was on January 25th, when the Covid-19 risk level was upgraded as "medium" by WHO. The other was on February 28th, when WHO raised the Covid-19 risk level to "very high". After March 23rd, the Covid-19 discussion intensity on Twitter dissipated gradually, and the Ratio index lowered to around 2% by the end of June 2020.

The global daily case rate (blue line) and mortality rate (red line) also changed over the six months, which ranged from 0.00 to 247.61 and 0.00 to 10.82 per ten million population respectively. Their average values were 83.99 and 4.09. Both daily case rate and mortality rate were low in January and February 2020, when Covid-19 cases were reported in only a few countries. Since the beginning of the pandemic-outbreak period, both indictors started to increase rapidly. The global daily case rate reached around 100 per ten million in late March and through April. The maximum global daily mortality rate was observed on April 17th, 2020, one month after the Twitter-derived public awareness reached the highest. The daily case rate increased gradually and peaked at the end of June, while the daily mortality rate declined progressively in May and stayed at around 6 per ten million in June 2020.

The global case fatality ratio (green line) showed a relatively smoother trend compared to daily case and mortality rate, ranging from 2.04% to 7.22%, with an average value of 4.95%. In late January, the case fatality ratio displayed a downward trend and reached the lowest value on February 5th. Then the case fatality ratio showed an upward trend and reached the maximum value on April 29th, 2020, around 37-48 days after the public awareness reached the highest on Twitter. Finally, the case fatality ratio gradually decreased to 4.87% in May and June.

Figure 4 describes the lag Pearson's correlations between the global daily Ratio index and case rate (a), mortality rate (b), and case fatality ratio (c). We set the maximum number of days' difference to 60. The largest correlation coefficients between Ratio and the Covid-19 mortality rate and case fatality ratio were 0.80 and 0.86, and the leading days were 22 and 38 days, respectively. No turning point (the point where the lag correlation coefficient starts to decline) was detected in the lag correlation between Ratio and case rate, and the maximum Pearson's correlation was 0.52.

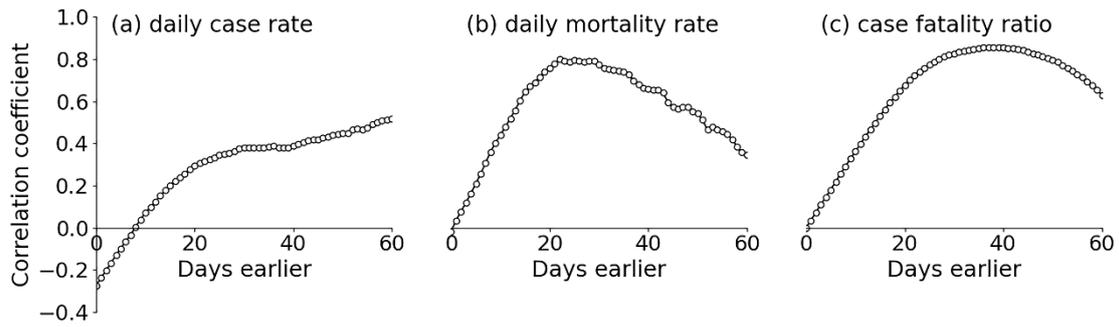

Figure 4. Lag Pearson correlation between global Ratio index and daily case rate (a), daily mortality rate (b), and case fatality ratio (c).

Three phenomena were observed from the global temporal trends. Initially, public awareness of Covid-19 reflected on Twitter significantly correlated with the daily mortality rate and case fatality ratio. The result is consistent with another analysis exploring the possibility of forecasting the Covid-19 outbreak in China from social media data, which found significant correlations between Sina Weibo (a Chinese social media platform for microblogging) index and confirmed/suspected Covid-19 cases (Li et al. 2020). It confirms that monitoring social media activities can provide an accurate and timely prediction about the global outbreak and progression of infectious diseases like Covid-19. We also found that the global public awareness peak on social media was 21-30 days (correlation > 0.77) and 35-42 days (correlation > 0.85) earlier than the maximum daily mortality rate and case fatality ratio, respectively. The lagged days between global public awareness changes and Covid-19 health impacts are longer than the lag effect estimations in other research. For example, Chunara et al. (2012) concluded that trends in the volume of informal sources related to Haitian Cholera, including the news media reports and Twitter postings, were two weeks earlier than official case data. The difference comes from three reasons. First, Covid-19 is a more destructive epidemic that affects the entire human being and lasts a more extended period than other infectious diseases. Second, the indicators selected to represent Covid-19 health impacts are inconsistent among various investigations. The previous research chose infections to describe the health impact. The detected lag effects were on mortality and case-fatality ratio, whose trends are usually delayed compared with the infection trend. Third, the keywords used for tweet collection in this study are different from the words used in other studies. Culotta (2010) and Iso et al. (2016) have proved that choosing different keywords to collect disease-related information on social media could result in uneven leading effects.

Furthermore, Twitter users' awareness in the pre-pandemic period was mainly media-driven. The discussion intensity peaks corresponded with WHO announcements, meaning people tweeted about Covid-19 when official agencies reported it on social media platforms. It demonstrates that providing official updates and reports is an effective approach to enhance public awareness in the pre-pandemic period. Finally, although Covid-19 caused proliferating morbidity and mortality worldwide after becoming a pandemic and mass media and agencies continued to report this event, such social attention and health impacts were not entirely reflected in public discussions of Twitter users globally. The global Ratio index decreased gradually after March 18$^{th}$, with no significant discussion peaks by the end of June 2020.

## 4.2 Global Linguistic Disparities

Figure 5 reveals the proportions of tweets and the Ratio index by language from January 1$^{st}$ to June 30$^{th}$, 2020, which uncovers the global linguistic disparities of public awareness toward Covid-19. The numbers of background and Covid-19 tweets used to calculate the Ratio index are listed in Appendix Table A2. The most frequent tweeting language was English, followed by Japanese, Spanish, Portuguese, and Thai, consisting of 33.53%, 18.21%, 8.57%, 7.72%, and 6.32% tweets. Other common languages consisting of over 2% global tweets were Arabic, Korean, Indonesian, Turkish, and French.

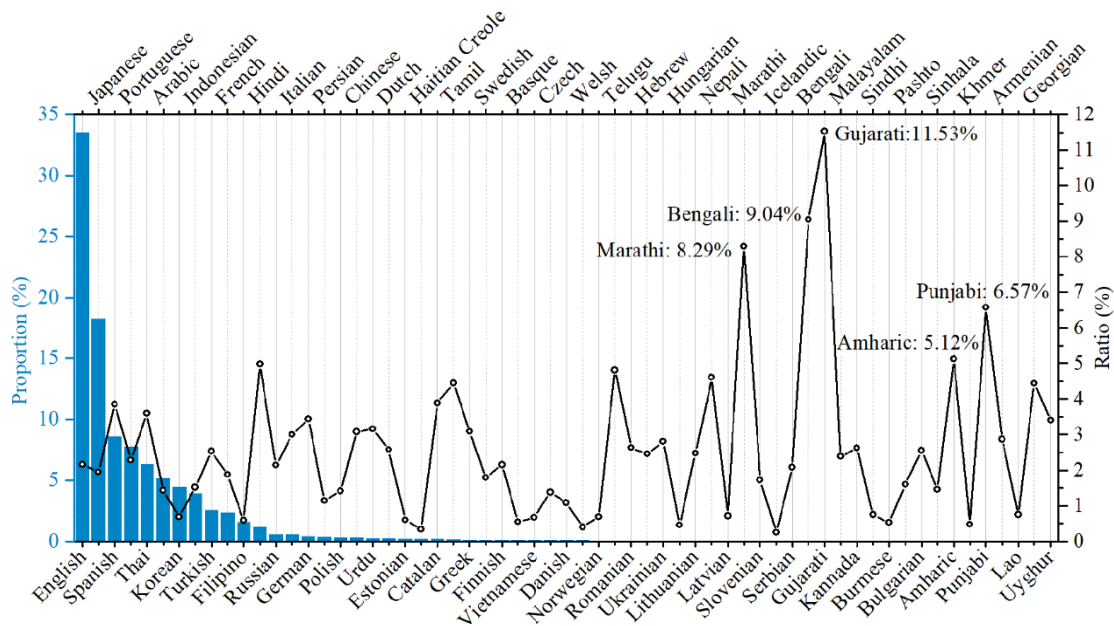

Figure 5. Proportions of Tweets and Ratio Index by languages.

The Ratio index by languages ranged from 0.24% to 11.53%, with an average value of 2.53% and a standard deviation of 2.17%. The highest Ratio index was in Gujarati language tweets. Users tweeting in Bengali, Marathi, Punjabi, and Amharic languages also showed high public awareness with Ratio index values of 9.04%, 8.29%, 6.57%, and 5.12%, respectively. The lowest Ratio index values were found in users tweeting in Icelandic, Haitian Creole, Welsh, Hungarian, and Khmer, and their values were 0.24%. 0.34%, 0.40%, 0.46%, and 0.48%, respectively.

The top four languages presenting the highest public awareness toward Covid-19 on Twitter are Gujarati, Bengali, Marathi, and Punjabi, which are the official languages in India. Bengali is also the most widely spoken language in Bangladesh. Both India and Bangladesh Governments practiced a nationwide lockdown at the end of March, 2020, and extended the lockdown several times in April and May, resulting in a two-month lockdown. Meanwhile, the new confirmed cases in Bangladesh grew by 1,155% in early April, the highest in Asia. Such societal and health impacts caused by the pandemic aroused public concerns and awareness about Covid-19 in India and Bangladesh, leading to more intense discussions on social media platforms like Twitter.

Figure 6 shows monthly variations of the Ratio index by the 20 most common languages on Twitter, indicating the temporal linguistic disparities of public awareness toward Covid-19. In each subplot, the bold green or blue line represents the monthly Ratio index in each language, while the monthly Ratio indexes in the other 19 languages are grey lines in the background. Numbers on the left top are the highest Ratio values in each language. The maximum monthly Ratio index was found in Hindi in April, and the value was 9.25%, followed by Thai with a value of 9.13% in March. Tweets in Filipino and Korean displayed the lowest maximum monthly Ratio index values of 1.24% and 2.29%.

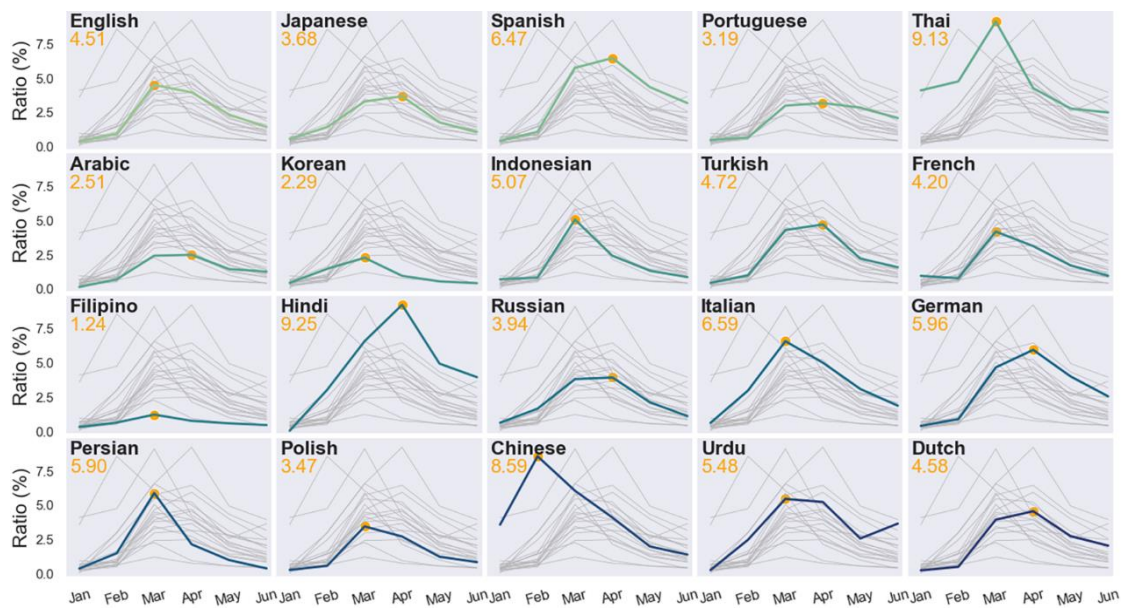

Figure 6. Monthly trends of the Ratio index by top 20 most popular languages on Twitter.

Most users grouped by tweeting languages showed the highest monthly Covid-19 awareness in March and April, except for users tweeting in Chinese which expressed the highest awareness in February 2020 with a value of 8.59%. This phenomenon matched the expectation, as the Covid-19 outbreak was reported first in China in January 2020. The Chinese government quickly responded and issued a series of policies and strategies to prevent the spread of the virus in January and February, such as Wuhan lockdown, isolation measures, and the construction of Fire God Mountain hospital, which raised public awareness among Chinese speakers worldwide. As Covid-19 became a pandemic since February 2020, the awareness among users speaking different languages began to increase.

It is worth mentioning that users tweeting in Thai also expressed high level of Covid-19 awareness in the pre-pandemic period. The following reasons could explain this observation. Thailand was the most popular traveling destination for residents of Wuhan. Thus, people living in Thailand were concerned that the frequent commute between Thailand and Wuhan might spread the virus from the reported outbreak center to Thailand. Meanwhile, Thailand confirmed the first Covid-19 case on January 13th, 2020, making it the first country other than China that had Covid-19 infected patients. Since then, the number of confirmed cases in Thailand grew to 25 on February 4th, more than any country other than China. Further, the government in Thailand was one of the very few governments that issued an early warning of Covid-19 in January 2020. On January 28th, their health minister claimed that the spread of the coronavirus could not be stopped. Both the health impacts caused by the pandemic and the government's immediate responses stimulated residents to discuss it on social media.

## 4.3 Country-Level Spatial-Temporal Patterns

The 29 countries with unique official/primary languages were included in the country-level spatial-temporal analysis. Figure 7 shows the spatial patterns of the Ratio index at the country level from January 1st to June 30th, 2020. The value range was from 0.24% to 4.62% and the global average was 2.12%. In general, Nepal and Thailand from southeastern Asia had the highest Twitter-derived public awareness of Covid-19. The public awareness on Twitter in the eastern part of Europe was also higher than the awareness in other countries. Iceland showed the lowest Ratio index. The Ratio index in the selected Asian countries ranged from 0.48% to 4.62% with a standard deviation of 1.31%, showing a larger disparity than the country-level Ratio index values in Europe (range: 0.24~3.89%, standard deviation: 0.98%).

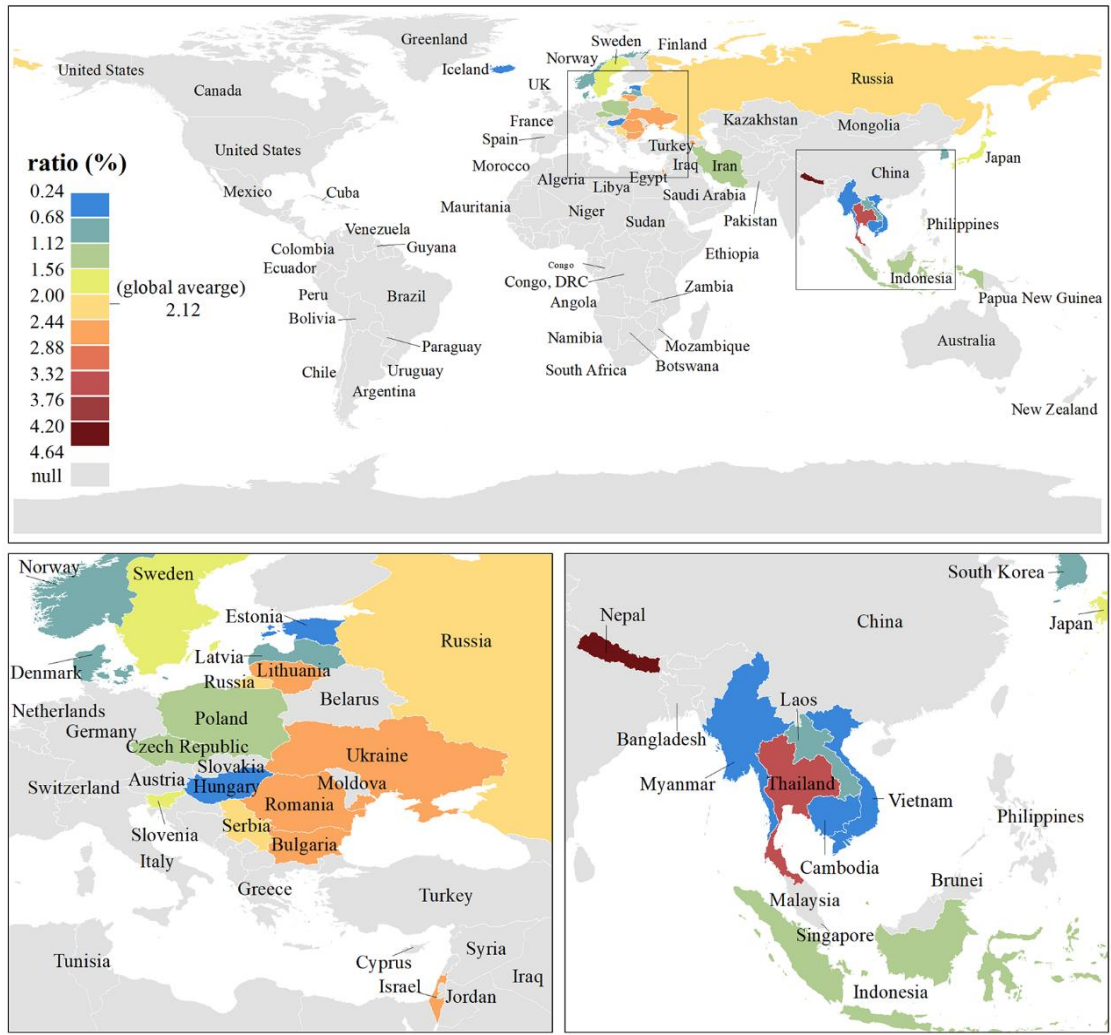

Figure 7. Global spatial patterns of the Ratio index at the country level from January 1st to June 30th, 2020.

Countries in Europe showed significant geographical discrepancies in Covid-19 awareness. Public awareness in the eastern part of Europe, including Russia, Armenia, Ukraine, Romania, Bulgaria, and Serbia, were generally higher than the awareness of countries in central Europe, including Norway, Denmark, Sweden, Estonia, Latvia, Lithuania, Poland, Czech Republic, Hungary, and Slovenia. The high public awareness on Twitter in the eastern part could be caused by two reasons. First, those countries were particularly vulnerable to the pandemic because their economies depended heavily on Western Europe, which was severely affected by disruptions in international production and transportation due to Covid-19 (Bohle 2021). Second,

Covid-19 caused more severe losses of working hours in the eastern part of Europe than in central Europe in 2020, based on the "Covid-19 and the world of work" report (International Labour Organization 2021). Therefore, residents in the eastern part of Europe were more concerned about Covid-19 and posted more discussions on social media platforms.

The monthly country-level Ratio index reveals detailed information about how public awareness toward Covid-19 on Twitter changed through different phases of the pandemic (Figure 8). The monthly Ratio index ranged from 0% to 13.98%. Similar to the observed global and by language patterns, Twitter users in most countries showed the highest awareness in March and April 2020, and expressed low awareness in the pre-pandemic period and in May and June 2020.

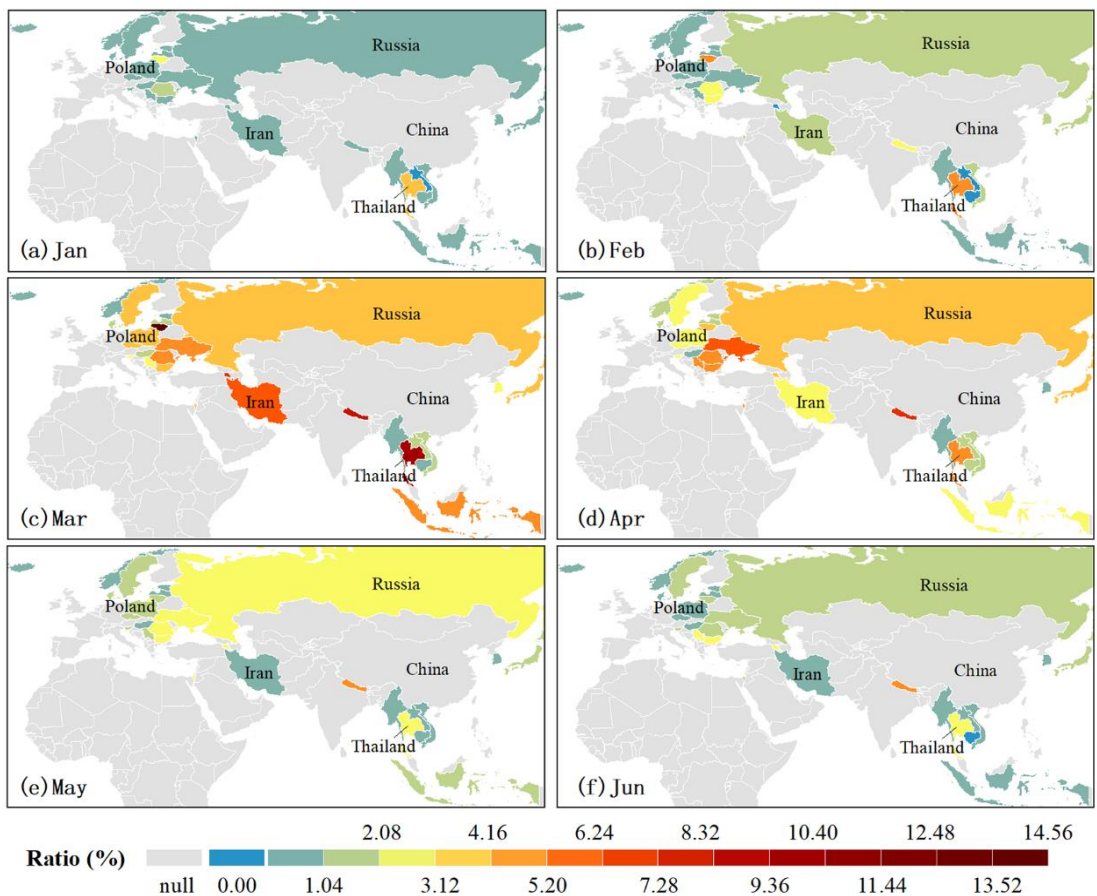

Figure 8. Spatial-temporal patterns of the monthly Ratio index at the country-level.

As mentioned in section 4.2, public awareness in Thailand was significantly higher than in other countries in January 2020. Lithuania, Thailand, Nepal, Bulgaria, and Romania showed higher values of monthly Ratio index in February than other countries, with values of 5.13%, 4.77%, 2.62%, 2.40%, and 2.38%, respectively. The highest monthly awareness on Twitter was in Lithuania in March. Although Lithuania had only three confirmed cases by March 12$^{th}$, the Lithuanian government cancelled all public indoor events of more than 100 attendees and closed all educational institutions (i.e., kindergartens, public schools, and universities) and entertainment places (i.e., museums, cinemas, and gyms) since that day. The shutdown had a massive impact on Lithuanians and intensified Covid-19-related discussion on Twitter. Israel presented the fastest growth rate of public awareness from 0.22% in January to 4.63% in April during the pandemic outbreak. The other countries with rapid awareness growth rates were Nepal, Iran, Serbia, and Bulgaria. In May and June 2020, only Nepal maintained a relatively high Covid-19 awareness.

**4.4 Covid-19 Public Awareness and Health Impacts**

Table 2 summarizes the maximum lag Pearson correlation coefficients (max r) and corresponding leading days between daily Ratio index and the three health impact indicators at the 29 countries within 60 days' lag effect. The daily mortality rate and case fatality ratio in Vietnam, Cambodia, and Laos were close to zero. Therefore, the lag Pearson correlations in the three countries could not be calculated and were labelled as 'null.' The 'nd' in table 2 indicates that a correlation turning point was not detected. Significant turning points were detected in 22, 21, and 11 countries regarding the lag Pearson correlations between the daily Ratio index and case rate, fatality rate, and case fatality ratio, respectively. The largest max r between daily Ratio index and the case rate was found in Thailand (0.83) with 10 leading days. The largest max r between the daily Ratio index and mortality rate and case fatality ratio were 0.80 and 0.74 in Sweden with 26-day and 39-day lag effects. Lag correlation coefficients between daily Ratio and case rate, mortality rate, and case fatality ratio were found the least significant in Iran, Bulgaria, and Armenia.

Table 2. Country level statistics of lag Pearson correlation between daily Ratio index and three health impact indicators

| Country | Case Rate | | Mortality Rate | | Case Fatality Ratio | | Country | Case Rate | | Mortality Rate | | Case Fatality Ratio | |
|---|---|---|---|---|---|---|---|---|---|---|---|---|---|
| | Max r | Lagged Days | Max r | Lagged Days | Max r | Lagged Days | | Max r | Lagged Days | Max r | Lagged Days | Max r | Lagged Days |
| Japan | 0.74** | 9 | 0.76** | 24 | nd | nd | Israel | 0.59** | 14 | 0.73** | 21 | nd | nd |
| Thailand | 0.83** | 10 | 0.76** | 18 | 0.34** | 39 | Ukraine | nd | nd | nd | nd | nd | nd |
| South Korea | 0.63** | 3 | 0.69** | 15 | nd | nd | Hungary | 0.50** | 11 | 0.40** | 25 | nd | nd |
| Indonesia | nd | nd | nd | nd | 0.60** | 23 | Lithuania | 0.52** | 23 | 0.48** | 28 | nd | nd |
| Russia | 0.81** | 50 | nd | nd | nd | nd | Nepal | nd | nd | nd | nd | nd | nd |
| Iran | 0.19 | 19 | 0.45** | 18 | nd | nd | Latvia | 0.46** | 19 | 0.45** | 40 | nd | nd |
| Poland | 0.55** | 36 | 0.73** | 42 | 0.70** | 49 | Slovenia | 0.53** | 16 | 0.56** | 29 | nd | nd |
| Estonia | 0.68** | 14 | 0.61** | 24 | nd | nd | Iceland | 0.38** | 2 | 0.47** | 32 | 0.52** | 32 |
| Andorra | 0.73** | 8 | 0.67** | 16 | nd | nd | Serbia | 0.57** | 29 | 0.50** | 16 | 0.47** | 17 |
| Sweden | nd | nd | 0.80** | 26 | 0.74** | 39 | Myanmar | nd | nd | 0.46** | 25 | 0.35** | 25 |
| Vietnam | 0.38** | 3 | null | null | null | null | Bulgaria | 0.30** | 47 | 0.28* | 46 | 0.38** | 4 |
| Czech Republic | 0.67** | 14 | 0.73** | 26 | nd | nd | Cambodia | nd | nd | null | null | null | null |
| Denmark | 0.58** | 14 | 0.54** | 21 | 0.46** | 37 | Armenia | nd | nd | nd | nd | 0.26* | 41 |
| Norway | 0.41** | 1 | 0.35** | 26 | 0.34** | 43 | Laos | 0.53** | 10 | null | null | null | null |
| Romania | 0.52** | 28 | 0.59** | 34 | nd | nd | Global | nd | nd | 0.8 | 22 | 0.86 | 38 |

* and ** indicates that the correlations of the Ratio index and health impact indicators are significant at p<0.01 and p<0.001 levels.

The daily Ratio index showed uneven abilities in predicting case rate, mortality rate, and case fatality ratio (Table 2). The average max r between Ratio and mortality rate was 0.57. It is higher than the averaged correlation coefficients between Ratio and case rate (0.55) and case fatality ratio (0.47). It demonstrates that the Ratio index has an overall better performance in predicting mortality rate in different countries during the Covid-19 outbreak. The average leading days between Ratio and mortality rate, case rate, and case fatality ratio were 17.27, 26.29, and 31.73, respectively.

The correlation coefficients between the daily Ration index and the 3 health impact indicators varied from country to country, leading to different levels of prediction performance of daily Ratio index in different countries. For example, daily Ratio index shows the best prediction ability for Covid-19 outbreak in eight countries, including Japan, Thailand, South Korea, Poland, Estonia, Andorra, Sweden, and the Czech Republic. The values of max r between the Ratio index and at least two health impact indicators exceeded 0.6 in the eight countries. On the contrary, none of the max r between Ratio and three health impact indicators were greater than 0.5 in Iran, Vietnam, Norway, Latvia, Myanmar, Bulgaria, and Armenia. No turning points in lag Pearson's correlations were detected in Ukraine and Nepal. The undesirable prediction performance of the daily Ration index is due to few Covid-19 cases in these countries during the time period which makes outbreak detection difficult, and insufficient Twitter data to evaluate the public awareness levels across time.

5. Discussion

5.1 Significant Implications

The completion of this research yields several significant implications. First, this research demonstrated that the language-country matching method could overcome the limitation of insufficient social media data with geotagged locations and efficiently geo-reference big Twitter data for large-scale analysis with an accuracy of 77%. The method works the best for countries with unique official languages, offering a novel and rapid channel observing and comparing social media activities, e.g., public awareness toward different topics or events, among those countries. The generated datasets provide baseline information on Covid-19 awareness globally and by language and country.

Second, this study found that the daily changes of public awareness reflected on Twitter had strong correlations with the daily mortality rate and case fatality ratio, verifying the near-real-time Twitter data's ability in predicting Covid-19 outbreak. This finding is consistent with prior investigations which applied social media to detect and forecast the outbreak of other infectious diseases, such as influenza (Signorini, Segre, and Polgreen 2011; Zadeh et al. 2019) and Cholera (Chunara, Andrews, and Brownstein 2012). Therefore, monitoring continuous social media activities could help forecast the future outbreaks of Covid-19 and its variants, and inform governments and communities to plan accordingly.

Finally, this study further proved that the prediction abilities of social media data for infectious disease outbreaks were distinct in different countries. It is in line with the analysis in Allen et al. (2016), which measured the correlation between Twitter rates and the official reports of influenza-like illness (ILI) in 31 major cities in the United States during the 2013–2014 flu season. Our results confirm that the spatial variability of disease detection performance based on the social media data needs to be considered and mitigated in future work to accurately predict the regional outbreak of infectious diseases and develop responding strategies.

## 5.2 Limitations and Future Research

A few limitations exist in this investigation and necessitate further research. First, the language-country matching method is unable to capture social media activities in countries speaking leading languages of international discourse, such as English and Spanish, or multilingual countries listed in Table A3. This limitation can be resolved by incorporating the locations obtained from the geotags, users' profiles, and tweet contents in future research.

Second, health impact indicators were derived from the cumulative Covid-19 cases and deaths dataset from CSSE. However, these data have inherent limitations due to the under-testing or under-reporting of cases. For example, inadequate diagnostic facilities especially in the initial outbreak stage of the pandemic, and delayed diagnosis could underestimate the case and death numbers. Such uncertainty in original health impact data can subject the prediction ability assessment of the Ratio index to unavoidable errors.

Third, many researchers have leveraged multiple big data sources, such as Facebook, Twitter, and Google trends, to track Covid-19 awareness. Data from social media platforms provide more about users' experience and feeling about the epidemic, while specific keywords analysis from search engines make it possible to distinguish what people want to know about Covid-19, such as infection symptoms or prevention methods. More studies assessing and comparing various aspects of pandemic awareness from multi-sourced big data should be considered.

Furthermore, although this study successfully tracked the spatial-temporal disparities of Covid-19 awareness, other dimensions of information reflected on Twitter, e.g., public attitudes and communications, were not considered. Differences in public attitudes shaped human behaviors during the outbreak, leading to unequal health impacts. Regions showing similar levels of Covid-19 awareness might present disparate attitudes, such as oppose, ignore, or support, to epidemic prevention measures and policies. Therefore, the public attitudes and communications toward the pandemic and relevant policies should be included in future research to further elucidate factors causing unequal Covid-19 health impacts.

Finally, future investigations could categorize Twitter data by different user groups, such as citizens, celebrities, scientists, health agencies, and governments, and by different pandemic phases, e.g., pre-pandemic preparedness, pandemic-outbreak response, and post-pandemic recovery. Examining the public awareness of the pandemic among diverse social groups in each phase could gain valuable information on how different opinion leaders on social media affect the citizens' Covid-19 awareness in different countries and communities and how changing public awareness impacts pandemic recovery.

## 6. Conclusion

This study analyzed the global Twitter data from January 1st to June 30th, 2020, aiming to answer two questions: what are the linguistic and geographical disparities of public awareness to Covid-19 on social media, and can the changing awareness predict the Covid-19 outbreak? The study established a social media data mining framework, which uses the Twitter-calculated Ratio index to quantify the disparities of pandemic awareness by different languages and at multiple spatial and temporal scales. The lag Pearson correlations between Ratio index and Covid-19 health impact indicators were calculated at the global scale and different countries, confirming significant associations and lag effects between public awareness and the Covid-19 outbreak.

There are valuable findings from this research. First, global public awareness of Covid-19 changed over the six months and reached the highest in middle March 2020. The major official pandemic-related announcements triggered the public awareness in the pre-pandemic period. The global public awareness dynamics on Twitter could accurately predict the mortality rate and case fatality ratio by 21-30 and 35-42 days ahead. Second, linguistic and geographical disparities of public awareness existed during the pre-pandemic and pandemic-outbreak periods. Users tweeting in India and Bangladesh's official or regional languages, e.g., Gujarati, Bengali, Marathi, and Punjabi, had higher public awareness toward Covid-19 due to the impacts caused by the nationwide lockdown in both countries. Users tweeting in Chinese were most concerned about the pandemic in February 2020, while the Covid-19 awareness of users tweeting in other languages was the highest in March and April of 2020. Asian countries had greater disparities in Covid-19 awareness than European countries, and Eastern Europe's public awareness was generally higher than that of central Europe. Third, the Ratio index performed the best in mortality rate prediction with an average of 17 leading days at the country level. The prediction ability of the Ratio index in Covid-19 health impacts varies from country to country, suggesting that in-depth research on how discrepancies in public awareness, together with diversified socioeconomic conditions, medical resources, and public attitudes in different countries, affect the Covid-19 morbidity and mortality are needed.

**Acknowledgements**

This article is based on work supported by the Texas A&M Institute of Data Science (TAMIDS) under the Data Resource Development Program. The statements, findings, and conclusions are those of the authors and do not necessarily reflect the views of the funding agency.

**Data Availability Statement**

The data used in this research were derived from the following resources available in the public domain: Internet Archive (https://archive.org/details/twitterstream?sort=-publicdate), United Nation (http://data.un.org/), and COVID-19 Data Repository by the Center for Systems Science and Engineering (CSSE) at Johns Hopkins University (https://github.com/CSSEGISandData/COVID-19).

**Disclosure statement**

No potential conflict of interest was reported by the authors.

**Appendix**

Table A1. Language-country matching list

| Unique Language | Countries | Unique Language | Countries |
|---|---|---|---|
| Japanese | Japan | Hebrew | Israel |
| Thai | Thailand | Ukrainian | Ukraine |
| Korean | South Korea | Hungarian | Hungary |
| Indonesian | Indonesia | Lithuanian | Lithuania |
| Russian | Russia | Nepali | Nepal |
| Persian | Iran | Latvian | Latvia |
| Polish | Poland | Slovenian | Slovenia |
| Estonian | Estonia | Icelandic | Iceland |
| Catalan | Andorra | Serbian | Serbia |
| Swedish | Sweden | Burmese | Myanmar |
| Vietnamese | Vietnam | Bulgarian | Bulgaria |
| Czech | Czech Republic | Khmer | Cambodia |
| Danish | Denmark | Armenian | Armenia |
| Norwegian | Norway | Lao | Laos |
| Romanian | Romania | | |

Table A2. Statistics of Twitter data by languages

| Language (*lang*) | No. Tweets | No. Covid | Language (*lang*) | No. Tweets | No. Covid |
|---|---|---|---|---|---|
| English (en) | 151,118,159 | 3,263,975 | Welsh (cy) | 229,040 | 906 |
| Japanese (ja) | 82,077,330 | 1,596,719 | Norwegian (no) | 213,472 | 1,457 |
| Spanish (es) | 38,626,504 | 1,487,397 | Telugu (te) | 199,270 | 9,596 |
| Portuguese (pt) | 34,807,023 | 794,356 | Romanian (ro) | 185,386 | 4,881 |
| Thai (th) | 28,482,604 | 1,022,338 | Hebrew (iw) | 168,289 | 4,137 |
| Arabic (ar) | 23,067,235 | 330,651 | Ukrainian (uk) | 164,352 | 4,603 |
| Korean (ko) | 19,983,745 | 137,642 | Hungarian (hu) | 161,497 | 742 |
| Indonesian (in) | 17,743,004 | 268,826 | Lithuanian (lt) | 160,940 | 3,982 |
| Turkish (tr) | 11,530,106 | 292,367 | Nepali (ne) | 160,768 | 7,422 |
| French (fr) | 10,500,930 | 196,851 | Latvian (lv) | 144,214 | 1,027 |
| Filipino (fil) | 6,942,291 | 40,800 | Marathi (mr) | 123,621 | 10,250 |
| Hindi (hi) | 5,228,420 | 260,161 | Slovenian (sl) | 101,204 | 1,741 |
| Russian (ru) | 2,401,764 | 51,324 | Icelandic (is) | 96,718 | 236 |
| Italian (it) | 2,375,338 | 71,253 | Serbian (sr) | 82,531 | 1,719 |
| German (de) | 1,765,954 | 60,630 | Bengali (bn) | 72,816 | 6,586 |
| Persian (fa) | 1,445,157 | 16,535 | Gujarati (gu) | 57,667 | 6,648 |
| Polish (pl) | 1,315,752 | 18,604 | Malayalam (ml) | 51,769 | 1,233 |
| Chinese (zh) | 1,240,427 | 38,168 | Kannada (kn) | 40,437 | 1,061 |
| Urdu (ur) | 1,160,934 | 36,706 | Sindhi (sd) | 38,725 | 287 |
| Dutch (nl) | 1,043,682 | 26,951 | Burmese (my) | 38,292 | 203 |
| Estonian (et) | 891,694 | 5,340 | Pashto (ps) | 30,585 | 487 |
| Haitian Creole (ht) | 825,277 | 2,793 | Bulgarian (bg) | 30,363 | 775 |
| Catalan (ca) | 805,375 | 31,324 | Sinhala (sd) | 15,719 | 228 |
| Tamil (ta) | 599,658 | 26,634 | Amharic (am) | 7,853 | 402 |
| Greek (el) | 406,347 | 12,533 | Khmer (km) | 7,155 | 34 |
| Swedish (sv) | 363,011 | 6,484 | Punjabi (pa) | 5,157 | 339 |
| Finnish (fi) | 302,715 | 6,513 | Armenian (hy) | 4,158 | 119 |
| Basque (eu) | 267,813 | 1,467 | Lao (lo) | 3,608 | 27 |
| Vietnamese (vi) | 258,881 | 1,721 | Georgian (ka) | 1,037 | 46 |
| Czech (cs) | 258,308 | 3,566 | Uyghur (ug) | 235 | 8 |
| Danish (da) | 232,562 | 2,502 | | | |

Table A3. Popular languages and their speaking countries

| Popular Language | Countries |
|---|---|
| English | United States/United Kingdom/Australia/Singapore/Canada/New Zealand/Swaziland/Zimbabwe/South Africa/ Pakistan/India/Philippines/Zambia/ Botswana/Nigeria/Seychelles/Ghana/Kenya/Liberia/Malawi/Mauritius/Namibia/Rwanda/Sierra Leone/Sudan/Tanzania/ South Sudan/St. Helena/The Gambia/Uganda/Guernsey/Isle of Man/Jersey/Anguilla/Antigua & Barbuda/Barbados/Belize/Bermuda/Dominica/Grenada/Jamaica/Montserrat/ The Bahamas/Guam/Kiribati/Niue/Papua New Guinea/Tonga/Tuvalu/ Guyana/Trinidad & Tobago/Cameroon/Lesotho/Ireland/Malta/Puerto Rico/Fiji/ Nauru/Palau/Tokelau/Vanuatu/Samoa |
| Spanish | Spain/Mexico/Colombia/Cuba/Chile/Peru/Venezuela/Costa Rica/Dominican Republic/El Salvador/Guatemala/Honduras/Nicaragua/Panama/Puerto Rico/Bolivia/Ecuador/Bonaire/Paraguay/Uruguay/Equatorial Guinea |
| Portuguese | Brazil/Angola/Cape Verde/Guinea-Bissau/Mozambique/Sao Tome & Principe/Portugal/Equatorial Guinea /Timor Leste |
| Arabic | Egypt/Libya/Morocco/Iraq/Syria/Sudan/Yemen/Oman/Algeria/Tunisia/Bahrain/ Jordan/Kuwait/Lebanon/ Mauritania/Qatar/Saudi Arabia/United Arab Emirates /Djibouti/Somalia/Comoros |
| Turkish | Turkey/Cyprus |
| French | France/Switzerland/Canada/Central African Republic/Congo/Congo, DRC/Benin/Burkina Faso/Cameroon/Cote d'Ivoire/Djibouti/Equatorial Guinea/ Gabon/Guinea/Mali/Niger/Senegal/Togo/Monaco/Haiti/St. Pierre & Miquelon/New Caledonia/Wallis & Futuna/Burundi/Jersey/Vanuatu/Chad/Comoros/Seychelles/Madagascar/Mauritius/Rwanda/Belgium/Luxembourg |
| Italian | Italy/Switzerland/San Marino/Vatican City |
| German | Austria/Germany/ Switzerland/Liechtenstein/Belgium/Luxembourg |
| Dutch | Belgium/Netherlands/Aruba/Suriname |
| Greek | Greece/Cyprus |
| Bengali | Bangladesh/India |